\documentclass[9pt,twocolumn,twoside]{osajnl}

\journal{josab}

\setboolean{shortarticle}{false} 

\usepackage{siunitx}
\usepackage{graphicx}
\usepackage{bm}

\usepackage{tikz}
\usetikzlibrary{patterns}

\title{Single-Frequency Microwave Imaging with Dynamic Metasurface Apertures}

\author[1,*]{Timothy Sleasman}
\author[1]{Michael Boyarsky}
\author[1]{Mohammadreza F. Imani}
\author[1,2]{Thomas Fromenteze}
\author[1]{Jonah N. Gollub}
\author[1]{David R. Smith}

\affil[1]{Center for Metamaterials and Integrate Plasmonics, Duke University, Department of Electrical and Computer Engineering, Durham, NC 27708, USA}
\affil[2]{XLIM-CNRS 123, Avenue Albert Thomas, 87060 Limoges, Cedex, France}

\affil[*]{Corresponding author: sleasmant@gmail.com}

\dates{Compiled \today}

\ociscodes{(090.2910) Holography, microwave; (110.1758) Computational imaging; (160.3918) Metamaterials.}

\begin{abstract}
Conventional microwave imaging schemes, enabled by the ubiquity of coherent sources and detectors, have traditionally relied on frequency bandwidth to retrieve range information, while using mechanical or electronic beamsteering to obtain cross-range information. This approach has resulted in complex and expensive hardware when extended to large-scale systems with ultrawide bandwidth. Relying on bandwidth can create difficulties in calibration, alignment, and imaging of dispersive objects. We present an alternative approach using electrically-large, dynamically reconfigurable, metasurface antennas that generate spatially-distinct radiation patterns as a function of tuning state. The metasurface antenna consists of a waveguide feeding an array of metamaterial radiators, each with properties that can be modified by applying a voltage to diodes integrated into the element. By deploying two of these apertures, one as the transmitter and one as the receiver, we realize sufficient spatial diversity to alleviate the dependence on frequency bandwidth and obtain both range and cross-range information using measurements at a single frequency. We experimentally demonstrate this proposal by using two one-dimensional dynamic metasurface apertures and reconstructing various two-dimensional scenes (range and cross-range). Furthermore, we modify a conventional reconstruction method---the range migration algorithm---to be compatible with such configurations, resulting in an imaging system that is efficient in both software and hardware. The imaging scheme presented in this paper has broad application to radio frequency imaging, including security screening, through-wall imaging, biomedical diagnostics, and synthetic aperture radar.
\end{abstract}

\setboolean{displaycopyright}{true}

\begin{document}

\maketitle

\section{Introduction}
\label{sec:intro}

Microwave imaging systems provide unique sensing capabilities that are advantageous for a variety of applications. Their features include the construction of three-dimensional (3D) images, the ability to penetrate optically-opaque materials, and the use of non-ionizing electromagnetic radiation---collective traits that are desirable in applications ranging from security screening to biomedical diagnostics \cite{Sheen2001, Meaney1998, Alvarez2015, Bond2003, Fear2002, Dehmollaian2008, Haynes2012, Ravan2010, Zhuge2012, Ahmed2011, Ahmed2012}. Most conventional microwave systems take the form of mechanically-scanned antennas or use electronic beamforming to retrieve scene content from backscatter measurements \cite{Sheen2001, Hansen2009, Soumekh1999, Steinberg1991}. Excellent imaging performance can be obtained from these systems, but these approaches often suffer from implementation drawbacks. Specifically, mechanically-scanned antennas tend to be slow and bulky, while electronic beamforming systems, such as phased arrays or electronically scanned antennas (ESAs) are complex, expensive, and often exhibit significant power draw \cite{phased_array_afford, fenn2000development}. 

\begin{figure}[t]
	\centering
	\includegraphics[width=3.4in]{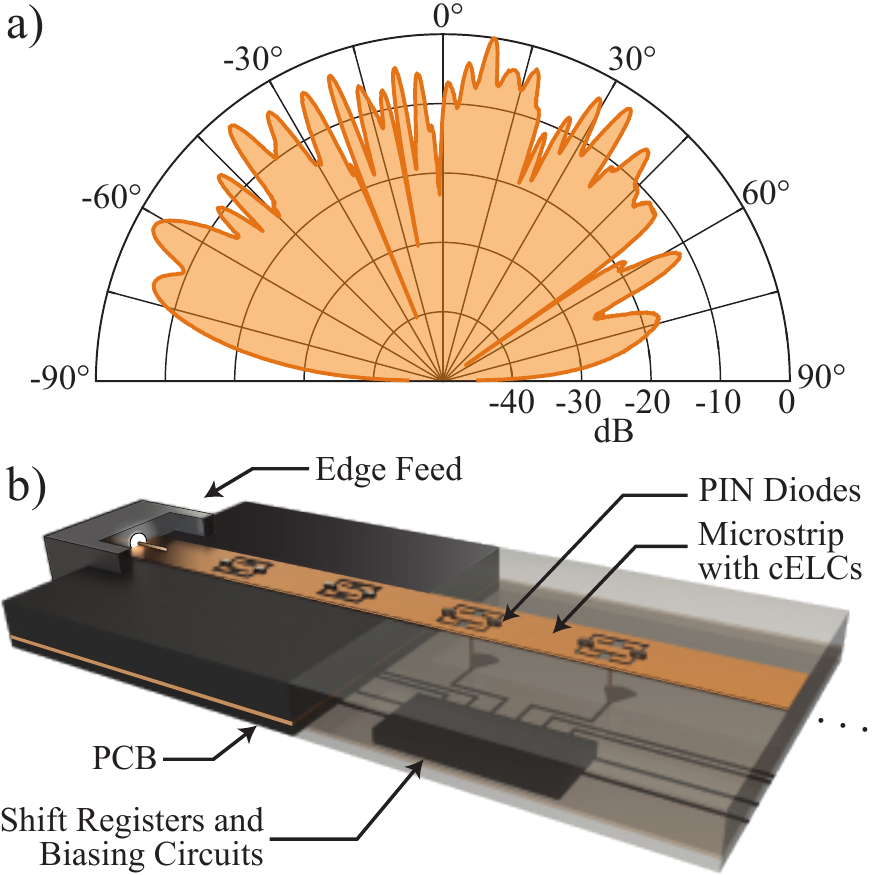}
	\caption{a) An example E-plane radiation pattern for the antenna used in this work and b) a schematic of the aperture, its RF feed, and the control circuitry. More details about the cell and the aperture biasing network can be found in \cite{Sleasman2016a,Sleasman2016d}.}
	\label{fig:apSchem}
\end{figure}

Since the radiation pattern from an antenna does not vary with range in the far-field, targets distant from a static antenna must be probed with a band of frequencies to resolve depth information in addition to cross-range information. Such scenarios are common, for example, in synthetic aperture radar imaging, which makes use of an antenna scanned over a large area \cite{Soumekh1999}. Imaging paradigms that require wide bandwidths complicate the radio frequency (RF) signal generation/detection hardware and necessitate allocation of large portions of the electromagnetic spectrum. When broadband operation is needed, the RF components---filters, power dividers, etc.---become more expensive and performance is sacrificed relative to narrowband components. Using a large frequency bandwidth also increases the possibility for interference with other electronic devices, hindering performance and complicating processing techniques. In addition, wideband systems have limited data acquisition rates (due to the extended dwell times needed to settle the phase locking circuitry). From an imaging perspective, objects exhibiting frequency dispersion (such as walls in through-wall imaging) are problematic for wideband systems \cite{Dehmollaian2008, Song2005}. From an implementation perspective, large bandwidths can also contribute to the complexity of system integration because they increase sensitivity to misalignment and necessitate calibration of the aperture/feed layers \cite{Odabasi2016}. Given these numerous challenges, reliance on frequency bandwidth has become the bottleneck of many microwave imaging systems. If the requirement of bandwidth can be removed, a microwave imaging platform may experience considerable benefits in many of these features.

The notion of simplifying hardware is in line with many emerging applications which require a capacity for real-time imaging within strict economic constraints. Efforts have recently been made to simplify the hardware required for high resolution imaging and instead rely more heavily on post-processing algorithms. In these approaches, termed \emph{computational imaging}, spatially-distinct waveforms can be used to interrogate a scene's spatial content, with computational techniques used to reconstruct the image. This idea, which transfers much of the burden of imaging from hardware to software, has been demonstrated across the electromagnetic spectrum as well as in acoustics \cite{Watts2014, Brady2009, Brady2013, F.Imani2016, Katz2014, Xie2015, LemoultMetaLens, Liutkus2014, Lemoult2011, Donoho2006, Eldar2012, Rossi2014}. Applying this idea to microwave imaging, the conventional dense antenna arrays (or mechanically-swept systems) can be replaced with large antennas that offer radiation pattern tailoring capabilities, e.g. metasurface antennas. Metasurface antennas often take the form of a waveguide or cavity structure that leaks out the contained wave through metamaterial elements. It has been shown that metasurface antennas can radiate spatially-diverse, frequency-indexed wavefronts in order to multiplex a scene's spatial information \cite{Hunt2013, Hunt2014, Gollub2017, Fromenteze2015, F.Imani2016}. In these demonstrations, post-processing decodes the measurements to resolve an image in all three dimensions. Since these systems rely on large antennas and collect information across a sizable surface in parallel \cite{Gollub2017}, the architecture is favorable compared to systems that are densely populated with independent antenna modules \cite{Gonzalez-Valdes2014}. Nonetheless, such systems remain particularly dependent on a wide bandwidth due to their use of frequency diversity to generate a sequence of spatially diverse radiation patterns.

Alternative efforts have focused on reducing the dependence of metasurface antennas on a wide bandwidth by leveraging electronic tuning \cite{Sleasman2016c, Sleasman2016d, SleasmanSubmittedforpublication-2016}. These \emph{dynamic metasurface antennas} are composed of individually addressable metamaterial elements whose resonance frequencies can be varied by application of an external voltage. A generic example of a dynamic metasurface antenna in the form of a microstrip waveguide, similar to the one used in this work, is shown in Fig. \ref{fig:apSchem}. By tuning the resonance frequency of each element---which influences the amplitude and phase of radiation emitted from that point---spatially-distinct radiation patterns are generated which multiplex a scene's cross-range content without requiring a large frequency bandwidth. An example of a spatially-diverse far-field pattern from a dynamic metasurface aperture (taken from the device used in this work) is shown in Fig. \ref{fig:apSchem}. Using a dynamic metasurface as a transmitter in conjunction with a simple probe as a receiver, high-fidelity imaging has been experimentally demonstrated using a small bandwidth, even down to a single frequency point \cite{Sleasman2016c, Sleasman2015, Sleasman2016d}. However, in these demonstrations range resolution was limited due to the small bandwidth of operation. In the extreme case of a single frequency point, imaging in the cross-range dimension was viable but range information became negligible to the point that resolving in range was considered infeasible.

The seemingly indispensable bandwidth requirement, considered necessary to resolve objects in the range direction, is not a fundamental limitation arising from the physics of microwave imaging systems. This can be understood by interpreting image formation through spatial frequency components (i.e. in $k$-space). In this framework, sampling each spatial frequency component ($k$-component) can yield information in the corresponding spatial direction. Analysis in $k$-space has been successfully employed in synthetic aperture radar (SAR) for decades to ensure images free of aliasing. More recently, it has been used to realize phaseless imaging and to implement massive, multi-static imaging systems for security screening \cite{Alvarez2015, Ahmed2011, Ahmed2012, Gonzalez-Valdes2014, Boyarsky2017_SAR}.

More generally, wavefronts propagating in all directions (sampling all spatial frequency components through the decomposed plane waves) can retrieve information along all directions, even when a single spectral frequency is used. The key to achieving this capability is to operate in the Fresnel zone of an electrically-large aperture, where electromagnetic fields exhibit variation along both the range and cross-range directions, allowing the spatial frequencies in both directions to be sampled with no bandwidth.

\begin{figure*}[t]
	\centering
	\includegraphics[width=6.0in]{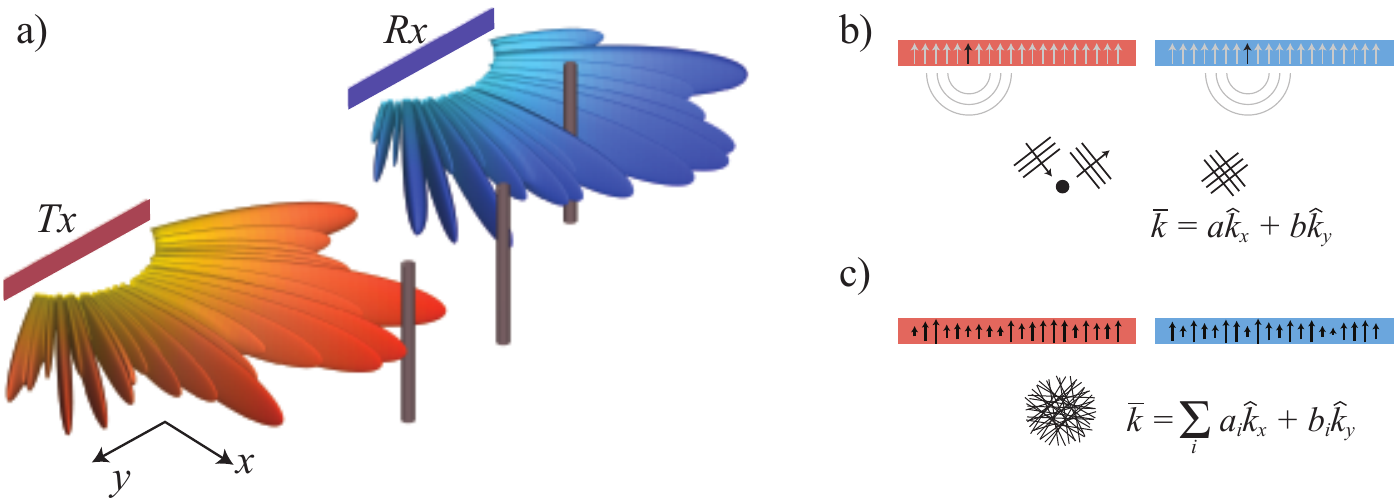}
	\caption{a) The overarching concept for computational imaging from an electrically-large dynamic aperture. The problem can be broken into b) a simple conceptual form but is more effectively implemented with a c) multiplexing aperture configuration. In the multiplexing version the sources across the entire aperture (black arrows) are simultaneously radiating and the signals are decoupled in post-processing.}
	\label{fig:concept}
\end{figure*}

To properly sample the required $k$-components at a single frequency, a set of relatively uncorrelated wavefronts must be generated spanning a wide range of incidence angles. This type of wavefront shaping has been explored from a $k$-space perspective in the optical regime, with the most famous examples being the early works on holography \cite{Gabor1948,Fienup1980}. In these works, a monochromatic source was sculpted into complex, volumetric shapes in the near-field of a recorded hologram. At microwave frequencies, however, generating waveforms to sample all $k$-components in an imaging application has not been straightforward. In a recent attempt a pair of simple antennas was mechanically scanned along separate, parallel paths and range/cross-range imaging was demonstrated using a single spectral frequency. This approach was successful, but proved to be prohibitively time consuming because every permutation of the two antennas' locations was needed \cite{Fromenteze2017}. In some quantitative imaging efforts, scanned antennas have been used in a tomographic setting to sample $k$-space and resolve objects in the along-range direction \cite{Song2005}. Other works have focused on single-frequency imaging, but these efforts have not been directed at obtaining range information and do not possess favorable form factors \cite{Li2016,Ravan2010,Sheen2001}. A more recent study has suggested using fields with orbital angular momentum (with different \emph{topological charges} creating different spiral patterns emanated from the antenna) to obtain range information with a single frequency, but the proposed structure is complicated and imaging performance has so far been limited to locating simple point scatterers \cite{Chen2017}.

Considering the results of $k$-space analysis and the ideas behind computational imaging, we investigate dynamic metasurface antennas as a simple and unique means for realizing single-frequency Fresnel-zone imaging. The proposed concept is illustrated in Fig. \ref{fig:concept} where an electrically-large, linear, transmit-receive aperture is formed by a pair of dynamic metasurfaces; one acting as transmitter and the other as receiver. To better visualize the unique features of this configuration, we have contrasted its sampling of $k$-space against the configuration used in \cite{Fromenteze2017}. Figure \ref{fig:concept}(b) demonstrates the configuration in \cite{Fromenteze2017} where two simple antennas were mechanically moved to sample all the possible $k$-components. This sampling is denoted by the rays from two arbitrary points on the transmit (Tx) and receive (Rx) antennas drawn according to the corresponding spatial frequencies at a point in the scene. Using such a configuration can only sample a small portion of $k$-space at each measurement (i.e. each synthetic aperture position). In Fig. \ref{fig:concept}(c) we show the $k$-space sampling of the configuration proposed in this paper. Since we are assuming Fresnel zone operation, $k$-components in 2D (both range and one cross-range dimension) can be probed from a 1D aperture radiating a series of distinct patterns at a single frequency. In other words, larger portions of the required $k$-space are sampled in each measurement.

In this paper, we experimentally demonstrate single-frequency imaging using dynamic metasurface antennas. We employ the 1D dynamic metasurface antenna previously reported in \cite{Sleasman2016d}, although the proposed concepts can be extended to other 1D and 2D architectures. The 1D dynamic metasurface considered here consists of an array of metamaterial resonators, with size and spacing that are both subwavelength. Each metamaterial element can be tuned to a \emph{radiating} or \emph{non-radiating} state (binary operation) using voltage-driven PIN diodes integrated into the metamaterial element design \cite{Sleasman2016a}. By tuning the elements in the transmitting (and receiving) aperture we modulate the radiation patterns illuminating the scene (and being collected), thus accessing large portions of $k$-space simultaneously (as demonstrated in Fig. \ref{fig:concept}(c)). The backscatter measurements collected in this manner are then post-processed to form 2D (range/cross-range) images. To facilitate the image formation, we modify the range migration algorithm (RMA) to be compatible with this specific metasurface architecture and mode of operation \cite{Fromenteze2016,Pulido-Mancera2016c}. Since this algorithm takes advantage of fast Fourier transform (FFT) techniques, it can reconstruct scenes at real-time rates \cite{Solimene2014}. Given that dynamic metasurface antennas are planar and easily-fabricated, the added benefit of single-frequency operation further simplifies the RF hardware requirements. The imaging system proposed in Fig. \ref{fig:concept}(a) therefore promises a fast, low-cost device for applications such as security screening, through-wall imaging, and biomedical diagnostics \cite{Bond2003, Fear2002, Dehmollaian2008, Sheen2001, Ahmed2012, Song2005}.

We begin by reviewing the underlying mechanisms of microwave and computational imaging. Next, we outline the specific aperture that has been implemented and detail its operation in a single-frequency imaging scheme. Experimental images are shown to highlight the performance and a point spread function analysis is carried out across the field of view \cite{Soumekh1999,Yurduseven2015}. We also demonstrate that single-frequency imaging is robust to practical misalignments---a property especially attractive in the emerging application of imaging from unmanned aerial/terrestrial vehicles with imprecise tracking capabilities \cite{watts2012unmanned}.

\section{Imaging Background}
\label{sec:background}

\subsection{Single-Frequency Imaging}

The range resolution of a microwave imaging system is typically set by the time domain resolution of the measured signal, equivalently stated in the frequency domain as \cite{Soumekh1994}

\begin{equation}
\delta_{\text{range}} = \frac{c}{2B}
\label{eq:rangeres}
\end{equation}

\begin{figure}[t]
	\centering
	\includegraphics[width=3.4in]{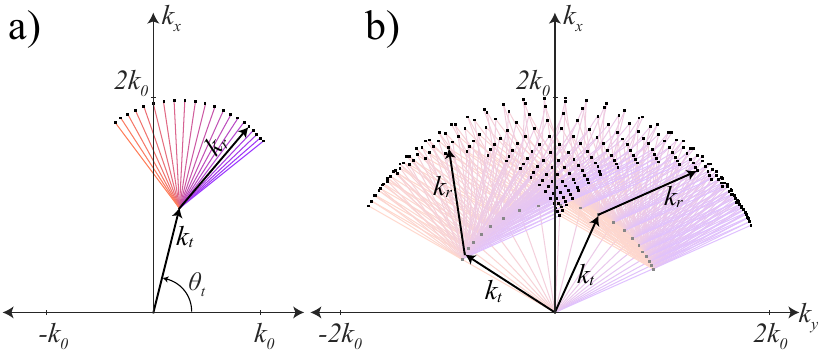}
	\caption{The $k$-space illumination at an off-centered location in a scene for the cases of a) SIMO and b) MIMO systems. Note that the $k$-space has significantly enhanced support in both $k_x$ and $k_y$ for the MIMO case.}
	\label{fig:SMIMO}
\end{figure}

\noindent where $c$ is the speed of light and $B$ is the frequency bandwidth. For a finite aperture with largest dimension $D$, the cross-range resolution, governed by diffraction, is given by \cite{Zhuge2012}

\begin{equation}
\delta_{\text{cross-range}} = \frac{R \lambda_0}{D}
\label{eq:crossrangeres}
\end{equation}

\noindent for a mean operating wavelength of $\lambda_0$ and a standoff distance $R$. For radar type schemes that rely on a frequency bandwidth to estimate range (or, equivalently, time-of-flight in the time domain), it is implied from (\ref{eq:rangeres}) that range resolution can only be improved by using a large frequency bandwidth.

In general, Eqs. (\ref{eq:rangeres}) and (\ref{eq:crossrangeres}) can be derived by analyzing the support in $k$-space. To better understand this relationship, we designate $x$ as the range direction and $y$ as the cross-range direction, as shown in Fig. \ref{fig:concept}. In both cases the resolution is given by \cite{Ahmed2011}

\begin{equation}
\delta_{x/y} = \frac{2\pi}{\Delta k_{x/y}} = \frac{2\pi}{k_{x/y,\text{max}}-k_{x/y,\text{min}}}.
\label{eq:delta_xy}
\end{equation}

\noindent Thus, to determine the range resolution we must identify the maximum and minimum values for $k_x$, which itself can be written as the sum of the transmitter (denoted by subscript $t$) and receiver (subscript $r$) components as

\begin{equation}
\begin{split}
	k_x & = k_{xt} +k_{xr} \\
        & = \sqrt{ k^2 - k_{yt}^2} + \sqrt{ k^2 - k_{yr}^2}
\end{split}
\label{eq:kxSum}
\end{equation}

%$k_{yt}\,{\approx}\,k_{yr}\,{\approx}\,0$
\noindent where we have used the dispersion relation $k_x\,{=}\,\sqrt{k^2 - k_y^2}$. When $x\,{\gg}\,y$ then $k_{yt}, k_{yr}\,{\ll}\,k$, which results in the bound $k_{x,\text{max}}\,{=}\,k_{xt,\text{max}}\,{+}\,k_{xr,\text{max}}\,{=}\,2 \pi f_{\text{max}}/c\,{+}\, 2 \pi f_{\text{max}}/c\,{=}\,4 \pi f_{\text{max}}/c$. This is a hard upper bound on what $k_{x,\text{max}}$ can be and it is true regardless of the system geometry. With the same $x\,{\gg}\,y$ assumption the minimum $k_x$ would similarly be $k_{x,\text{min}}\,{=}\,4 \pi f_{\text{min}}/c$. The difference can be used to derive $\delta_x\,{=}\,c/2B$ from (\ref{eq:delta_xy}), which would suggest that $B\,{>}\,0$ is necessary for range resolution. However, when $x\,{\approx}\,y$ the $k_{yr}$ and $k_{yt}$ terms become non-negligible and $k_x$ can be significantly smaller that $4 \pi f_{\text{min}}/c$. In this case, the $k_{x,\text{min}}$ becomes heavily dependent on the system's geometry.

When operating in the Fresnel zone of the imaging aperture, the geometry will have significant effects on the generated wavefronts. For example, a transmitting source located at the edge of the aperture will illuminate points in the scene with plane waves at oblique angles, sampling both the $k_x$- and $k_y$-components. In other words, different locations along the aperture generate different plane waves, spanning both $k_x$ and $k_y$ based purely on spatial diversity. This can be seen when examining the case of a multiple-input multiple-output (MIMO) system and tracking the resultant $k$-vectors, as seen in Fig. \ref{fig:SMIMO}. This case is contrasted with the single-input multiple-output (SIMO) version in Fig. \ref{fig:SMIMO}(a), where it is seen that the enhancement in spatial diversity delivered by using two large apertures is substantial. Having a single transmitter (or, equivalently, receiver) only allows for an arc in $k$-space to be sampled with a single frequency, whereas the pairwise combinations in a MIMO array can sample an area in $k$-space. Even though the configuration in Fig. \ref{fig:concept} can be viewed as a virtual MIMO system, it is important to note that the transmit aperture is only excited by a \emph{single} source and the received signal is only sampled by a \emph{single} detector.

In this imaging strategy, the $k$-space support becomes spatially-variant because different portions of the scene are probed by different sets of $k$-components. This results in resolution that depends on the position in the scene, as we will show in Section \ref{sec:exp}\ref{ssec:PSF}. The most important conclusion from this analysis is that, since the transmit-receive aperture samples both $k_x$- and $k_y$-components, it is possible to retrieve range information even at a single frequency, i.e. without any spectral bandwidth---contrary to Eq. (\ref{eq:rangeres}).

The analysis above sets guidelines for the imaging apertures and the effective sources across it. Specifically, two conditions should be met in order to sample the $k_x$- and $k_y$-components at a single spectral frequency: (i) an electrically-large transmit-receive aperture should be realized and (ii) different locations on this aperture must be sampled independently. A proof-of-concept effort was recently conducted to this end in which dipole antennas were translated to create an aperture satisfying these conditions \cite{Fromenteze2017}. These antennas were mechanically scanned through all locations on each aperture. (For each location on the synthetic Tx aperture, the receiver was scanned along the whole synthetic Rx aperture.) This experiment demonstrated the possibility to retrieve high-fidelity range/cross-range information using a single frequency. However, relying on mechanically-scanned synthesized apertures resulted in an impractical hardware implementation whose acquisition time was extremely slow. An alternative would be to use an array of independent transmitters and receivers, but this approach scales to a generally complex and expensive system. Instead, computational imaging has been studied as a paradigm which offers greater simplicity and more flexibility in hardware. %Instead, an imaging platform with greater simplicity and more flexibility has been sought after.

\subsection{Computational Imaging}

Interest in computational imaging has been on the rise due to the promise of simpler and less expensive hardware \cite{Brady2009a, Hunt2013}. The overarching concept behind computational imaging is to alleviate the burden of complex hardware and instead rely on post-processing to reconstruct an image. This often involves multiplexing data within the physical layer and extracting the important portions in software post-processing to build an image \cite{Brady2009a}. At optical and terahertz frequencies, several computational imaging systems have been proposed and experimentally demonstrated, including multiply-scattering media, random-lens arrays, and metamaterial spatial light modulators \cite{Brady2009,Watts2014,Fergus2006,Katz2014}. In the microwave regime, various metasurface designs have proven particularly useful for crafting apertures well-suited for computational imaging systems \cite{Hunt2013,Li2016,Hunt2014,Gollub2017}.

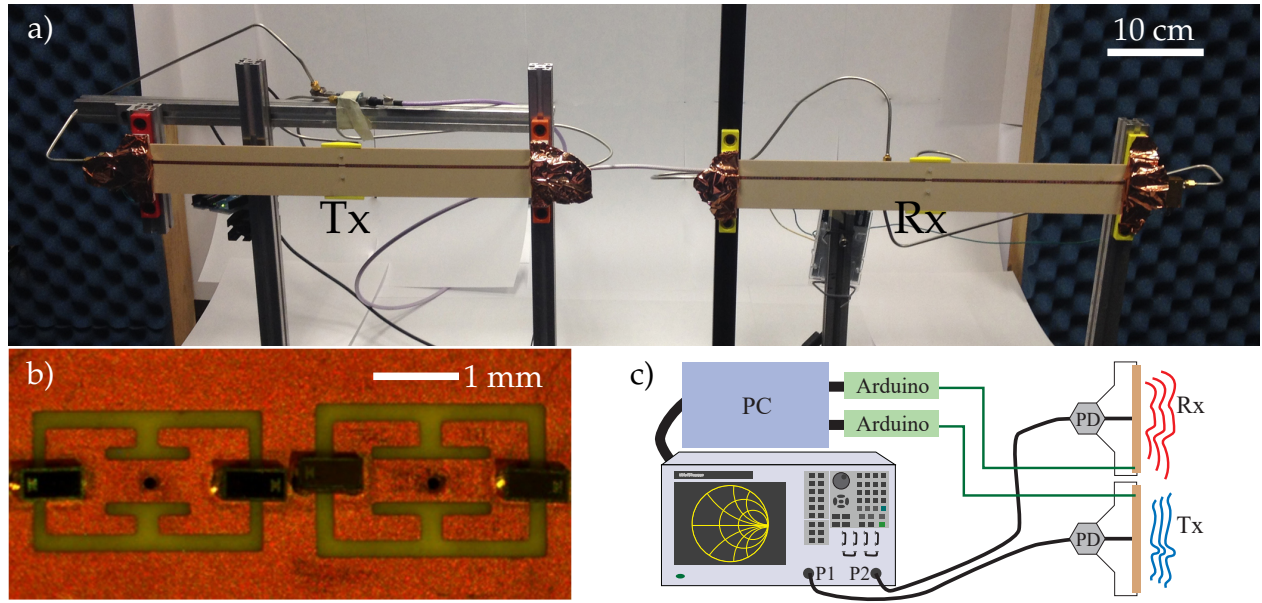
\begin{figure*}[t]
\centering

\begin{tikzpicture}

\node[anchor=south west,inner sep=0] (image) at (0.06,3.39) {\includegraphics[width=6.55in]{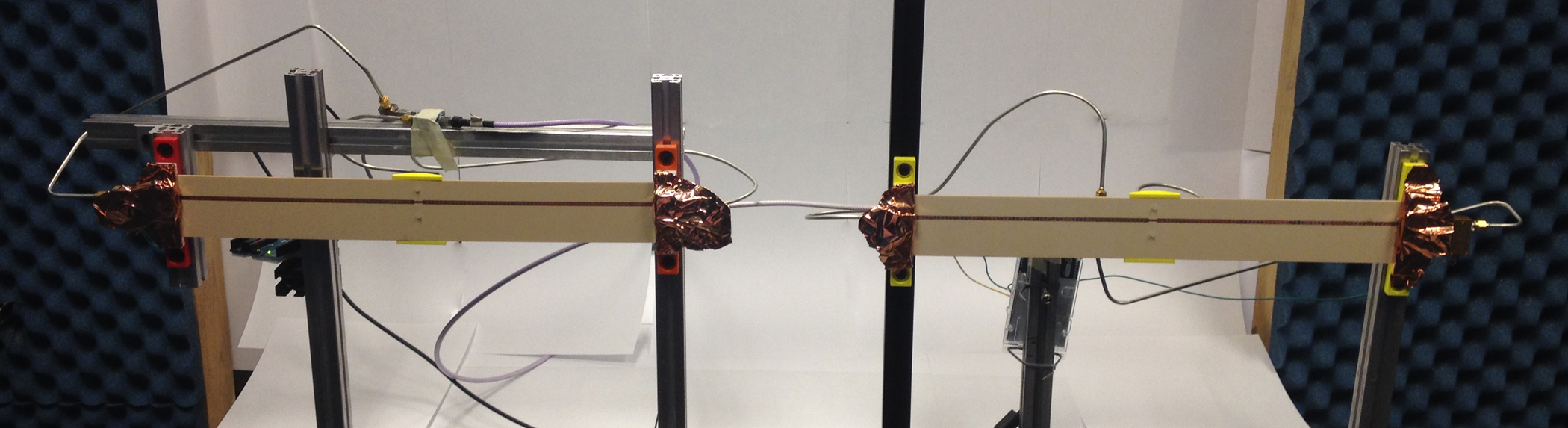}};
\draw[white,line width=1mm] (14.68,7.3) -- (15.96,7.3);
\node at (15.3,7.6) {\LARGE {\color{white} 10 cm}};
\node at (0.5,7.6) {\LARGE {\color{white} a)}};
\node at (4.55,5.1) {\Huge {\color{black} Tx}};
\node at (12.2,5.1) {\Huge {\color{black} Rx}};

\node[anchor=south west,inner sep=0] (image) at (0.06,0) {\includegraphics[width=2.95in]{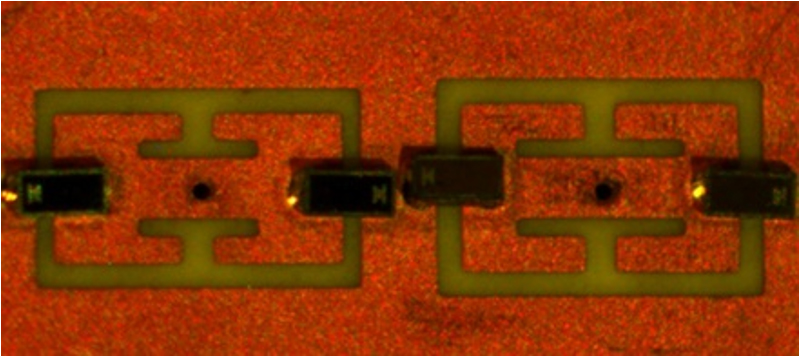}};
\draw[white,line width=1mm] (4.92,2.99) -- (6.06,2.99);
\node at (6.65,2.99) {\LARGE {\color{white} 1 mm}};
\node at (0.5,2.99) {\LARGE {\color{white} b)}};

\node[anchor=south west,inner sep=0] (image) at (8.5,-0.1) {\includegraphics[width=2.95in]{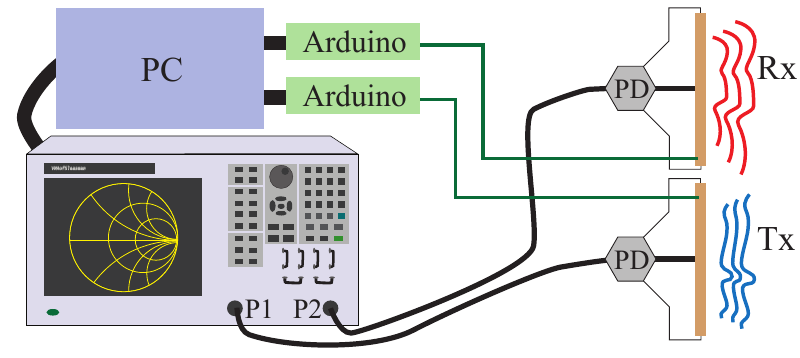}};
\node at (8.5,2.99) {\LARGE c)};

\end{tikzpicture}

\caption{a) The experimental setup for the single-frequency imaging, b) a close up of the metamaterial elements, and c) a schematic of the system configuration with Arudinos, feed lines, and power dividers (PDs).}
\label{fig:setup}
\end{figure*}

Metasurface apertures, which consist of waveguides loaded with numerous subwavelength elements, have been an architecture of much interest. They allow flexibility in design and also possess a key advantage in that the feed is directly integrated within the radiative layer (in contrast to a reflect array or a transmission mask \cite{Li2016,Watts2014}). Due to the subwavelength spacing of the individual elements and the large overall aperture size, a vast quantity of radiation patterns can be generated with significant spatial variation. These diverse patterns are vital to the computational imaging paradigm because they illuminate a large portion of the scene with complex field structures (as exemplified in Figs. \ref{fig:apSchem}(a) and \ref{fig:concept}(a)), allowing for the scene's spatial content to be encoded into a series of backscatter measurements \cite{Hunt2013,Hunt2014}. 

In order to alleviate the dependence on frequency diversity, dynamically-tunable metasurface antennas have been pursued \cite{Sleasman2015,Sleasman2016c,Sleasman2016d}. These antennas have tuning components, embedded into either the radiating elements \cite{Sleasman2016a} or the feed structure \cite{Sleasman2016c}, which enable the generation of diverse patterns as a function of externally applied voltage rather than driving frequency. In this work, we utilize the 1D dynamic metasurface pictured in Fig. \ref{fig:setup}. In this aperture, which is further described in Section \ref{sec:system}, each metamaterial element is loaded with a tunable component and is independently addressable via an applied voltage. This allows for modulation of the radiation as depicted in Fig. \ref{fig:concept}(c), even with a single frequency.

In \cite{Sleasman2016d}, this dynamic metasurface antenna was used as the transmitter of an imaging system while several dipole antennas were used as receivers. This configuration demonstrated high-quality imaging with a much smaller bandwidth than would be required for a frequency-diverse system \cite{Gollub2017}. The role of the dynamic metasurface was to generate spatially-distinct radiation patterns to resolve objects in the cross-range direction. Resolving objects in the range direction was still primarily accomplished using a finite bandwidth. However, the aperture from \cite{Sleasman2016d} can be adapted for multiplexing a 2D scene's spatial content (range and cross-range directions) with a single frequency. We use two apertures (one as a transmitter and another as a receiver) to cover a large area and we sample the spatial points along the apertures through linear combinations of the constituting elements. The dynamic metasurface can thus provide a practical alternative to the brute force bistatic SAR method utilized in \cite{Fromenteze2017}. In other words, instead of probing the scene from each location on the imaging aperture independently (as is the case in conventional schemes \cite{Sheen2001,Fromenteze2017} and is schematically shown in Fig. \ref{fig:concept}(b)), the signal is transmitted from and received at many locations on the aperture simultaneously, as depicted in Fig. \ref{fig:concept}(c). In a $k$-space interpretation, it can be said that the dynamic metasurface samples many $k$-components simultaneously, which are then de-embedded in post-processing. To better illustrate this concept, we review the operation of the dynamic metasurface and describe the imaging process in the next section.

\section{Single-Frequency Computational Imaging System}
\label{sec:system}

\subsection{Dynamic Metasurfaces}

The dynamic metasurface used in this paper is extensively examined in \cite{Sleasman2016d}. In this subsection, we briefly review its properties and describe its operation. This antenna is depicted in Fig. \ref{fig:setup} and consists of a $40$-cm-long microstrip line that excites 112 metamaterial elements. The metamaterial elements are complementary electric-LC (cELC) resonators \cite{Hand2008,Chen2007}, with every other element having resonance frequencies \SI{17.8}{\giga\hertz} and \SI{19.2}{\giga\hertz} (covering the lower part of K band). Each element is loaded with two PIN diodes (MACOM MADP-000907-14020W) and is individually addressed using an Arduino microcontroller which applies a DC voltage bias. The microstrip line is excited through a center feed (which generates two counter-propagating waves) and two edge feeds (each creating a single propagating wave) to ensure ample signal stimulates all of the elements. The edge feed can be seen in Fig. \ref{fig:apSchem} and the connectors for the end launch and the central launch are Southwest Microwave 292-04A-5 and 1012-23SF, respectively.

The signal injected into the aperture is distributed through power dividers and carried by the waveguide, but the structure is effectively a single port device due to the fact that a single RF signal is inserted/collected. The combination of the transmitter and receiver thus creates a single-input single-output (SISO) system, but emulates the performance of a MIMO system because tuning the elements allows us to multiplex the transmitted/received signal into a single feed. In this sense, the device retains simplicity in RF circuitry (only needing one feed, and allowing the RF components to be designed around a single frequency), but with the addition of the DC tuning circuitry. This DC circuitry, composed of shift registers, resistors, and diodes is vastly simpler than the design that goes into creating a broadband RF source/detector circuit.

The resulting radiation pattern generated from this single-port antenna is the superposition of the contributions from each constituting element. By turning each element \emph{on} or \emph{off} using the PIN diodes, the radiation pattern can be modulated without active phase shifters or moving parts. By leveraging the resonant behavior of the metamaterial elements, we can change the phase and magnitude of each radiator and manipulate the radiation pattern. The phase and amplitude of each metamaterial element cannot be controlled independently, but rather are tied together through the Lorentzian form of the element's resonance \cite{SmithLorentz}. While this connection places a limitation on the available field patterns, there is nevertheless enormous flexibility in possible patterns that can be exercised in this otherwise very inexpensive device. For example, in \cite{SleasmanSubmittedforpublication-2016} this structure was experimentally demonstrated to generate directive or diverse patterns depending on the applied tuning state. %The dynamic metasurface is planar, simply fabricated with printed circuit boards, and has an immense design space which can be explored for application-specific optimization \cite{Sleasman2015, Sleasman2016d}.

It is important to emphasize that this structure, while sharing some similarities with well-established leaky-wave antennas (LWAs), is distinct in both its architecture and operation \cite{Oliner1993, Grbic2002, Lim2004, martinez2013holographic, maci2015}. Both structures consist of a waveguide whose guided mode is gradually leaked through many small openings. However, leaky-wave antennas are designed such that the leaked radiation from each opening has a specific phase relative to the other openings, thus forming a directive beam which steers as a function of input frequency \cite{Oliner1993}. This is different from the dynamic metasurface, where the metamaterial radiators are placed at subwavelength distances apart and are not located/actuated to have a specific phase profile.

It should also be noted that the dynamic metasurface is different from the microwave camera described recently in \cite{Ghasr2012, Baumgartner2015}. While both structures use diodes to turn resonant radiators \emph{on}/\emph{off}, the microwave camera is designed such that only one element is radiating at a time and it has a complex corporate feed. This is quite different from a dynamic metasurface, where the radiation pattern is the superposition of several elements distributed over an electrically-large aperture. Phased arrays can also be compared to the dynamic metasurface antenna, but these structures often have phase shifters and amplifiers (and sometimes a full radio module) behind each element \cite{Hansen2009,fenn2000development}. While phased arrays exercise total control over the radiated fields, these structures are significantly more complicated than our dynamic aperture which has a single feed and no active components.
%It should also be noted that LWAs usually rely on frequency variation to modulate their patterns and that the microwave camera has a complex infrastructure with phase shifters.

\subsection{Imaging Setup}
\label{ssec:setup}

The imaging setup used in this work consists of two identical dynamic metasurfaces placed side by side with \SI{20}{\centi\meter} in between (edge-to-edge), as shown in Fig. \ref{fig:setup}. One dynamic metasurface is connected to Port 1 of an Agilent E8364c vector network analyzer (VNA) and acts as the transmitter, while the other is connected to Port 2 of the VNA and acts as the receiver. The RF signal on each side is distributed with a power divider to excite all three feed points, simultaneously generating four traveling waves with equal magnitude. Each dynamic metasurface is addressed separately using distinct Arduino microcontrollers. The DC control signals sent from the Arduinos are distributed to each independent element through a series of shift registers, allowing us to turn \emph{on}/\emph{off} combinations of elements at any time. The microcontrollers and VNA are controlled via MATLAB. Figure \ref{fig:setup}(c) shows a schematic of the control circuitry and the RF paths.

To characterize one of the dynamic metasurfaces, its near-field is scanned for different tuning states using a reference antenna \cite{Yaghjian1986}. The scanned data is then used along with the surface equivalence principle to represent the aperture as a set of discrete, effective sources. These sources can then be propagated to arbitrary locations in the scene, a process completed for both the transmit aperture and the receive aperture. The subwavelength metamaterial elements have mutual coupling which affects the overall radiation pattern, but these contributions (as well as any fabrication tolerances) are all captured in this experimental characterization and thus do not pose a problem.

During the imaging process the VNA is set to measure a single frequency, \SI{19.03}{\giga\hertz}, which provides ample SNR and diverse patterns from the aperture \cite{Sleasman2016d}. The transmitter's microcontroller enforces a tuning state $a_1$ and then the receiver's microcontroller sweeps through all the characterized tuning states $b_q$, $q=1,\ldots,Q$. The transmitter then moves on to the next tuning state and repeats, continuing this process through $a_p$, $p=1,\ldots,P$ until reaching the final state $a_P$. The raw data, a single complex number for each tuning state pair, is inserted into a [$Q \times P$] matrix $\mathbf{G}$. For the following experiments, we will use $Q=P$ such that the total number of measurements is $P^2$---e.g. for $50$ transmitter tuning states and $50$ receiver tuning states, $2500$ measurements will be taken at a single spectral frequency. It should be noted that the amount of information that a finite aperture can probe is inherently limited by its physical size and the operating frequency. The \emph{space-bandwidth product} imposes a physical limit on the number of independent free fields that can be generated from any aperture \cite{Piestun2000,Lohmann1996}. We have examined this consideration in previous works \cite{Sleasman2016d}, but will generally use a larger number of tuning states than is strictly necessary ($P\,{=}\,100$, compared to the space-bandwidth product, which is $50$ based on the antenna's size) to overcome potential correlation among the random patterns.

For each tuning state, a different combination of $30$ elements (out of the $112$) is chosen at random and set to the \emph{on} state. Randomness is chosen to ensure low correlation among the radiated patterns. Though a more deliberate set of \emph{on}/\emph{off} elements may give a minor enhancement in performance, the random selection process has been shown to provide ample diversity in the patterns \cite{Sleasman2016d}. The choice of $30$ \emph{on} elements is based on the analysis completed in \cite{Sleasman2016d}, which found that this quantity ensures that the guided mode is not depleted too quickly and that sufficient power is radiated. It is worth emphasizing that this process is purely electronic and could be accomplished in real-time by using a FPGA in conjunction with a customized RF source/detector.

\subsection{Image Reconstruction}
\label{ssec:recon}

In this paper, we use two different approaches to reconstruct images from the raw data. In the first approach, a sensing matrix, $\mathbf{H}$, is calculated based on the fields projected from the Tx and Rx antennas. By formulating the forward model of the Tx/Rx fields with the scene’s reflectivity, $\bm{\sigma}$, the set of measurements, $\mathbf{g}$, can be represented as

\begin{equation}
\mathbf{g} = \mathbf{H} \bm{\sigma}
\label{eq:gHf}
\end{equation}

\noindent where the quantities are discretized based on the diffraction-limit \cite{Lipworth2013,Lipworth2015}. The measurement column vector $\mathbf{g}$ is composed of the entries of the matrix $\mathbf{G}$ with order given by the indexing of the $(a_p, b_q)$ states. Assuming the Born approximation (weak scatterers) \cite{Lin1990}, entries of $\mathbf{H}$ are written as the dot product

\begin{equation}
\mathbf{H}_{ij} = \overline{E}_{Tx}^i (\overline{r}_j) \cdot \overline{E}_{Rx}^i (\overline{r}_j).
\label{eq:hexpand}
\end{equation}

\noindent The quantities $\overline{E}_{Tx}^i$ and $\overline{E}_{Rx}^i$ are, respectively, the electric fields of the Tx and Rx dynamic metasurface apertures for the $i$th tuning state combination. With the two apertures, this index is essentially counting the pairs $(a_1, b_1), (a_1, b_2), \ldots, (a_1, b_P), (a_2, b_1), \ldots, (a_P, b_P)$. The $\overline{r}_j$'s are the positions in the region of interest, resulting in each row of the $\mathbf{H}$ matrix representing the interrogating fields for all locations in the scene for a given tuning state, $(a_p, b_q)$. 

To reconstruct the scene reflectivity, Eq. (\ref{eq:gHf}) must be inverted to find an estimate of the scene $\hat{\bm{\sigma}}$. Since $\mathbf{H}$ is generally ill-conditioned, this process must be completed with computational techniques. Here, we use the iterative, least squares solver GMRES to estimate the scene reflectivity map \cite{Saad1986}.

While the method above is intuitive, it may require significant resources to calculate and store $\mathbf{H}$. In the system analyzed here, this issue does not cause practical difficulty; however, such limitations become problematic as the system scales up. A more efficient approach is to use range migration techniques, which employ fast Fourier transforms to rapidly invert a signal represented in $k$-space \cite{Soumekh1999, Stolt1978, Solimene2014}. The conventional range migration algorithm (RMA) assumes simple, dipolar antennas with well-defined phase centers---a condition not met by our dynamic metasurface aperture \cite{Lopez-Sanchez2000}. This issue has recently been addressed in \cite{Zhuge2012} and a pre-processing technique was introduced in \cite{Pulido-Mancera2016c,Fromenteze2016} to make the RMA and dynamic metasurfaces compatible. In the following, we will review this pre-processing step and then recap the salient portions of the RMA before adapting this method to the single frequency case used here.

The required additional step lies in transforming the data obtained by a complex radiating aperture into independent spatial measurements \cite{Pulido-Mancera2016c, Fromenteze2016}. In this case, where two distinct metasurface antennas are used as the transmitter and receiver, the pre-processing step must be applied independently for each metasurface antenna. For a single tuning state, the effective source distribution on the aperture plane can be represented as a [$1{\times}N$] row vector with the entries being the discretized sources. By cycling through $P$ tuning states, the effective sources can be stacked to obtain the [$P{\times}N$] matrix $\bm{\Phi}$, which represents the linear combinations that make up the radiating aperture. An inverse (or pseudo-inverse when $P \neq N$) is then used to diagonalize the matrix of measurements and find the contributions for each source location. The procedure is carried out through a singular value decomposition (SVD) with a truncation threshold, the latter technique eliminating subspaces that grow toward infinity and corrupt the data during inversion. The transformed signal $\mathbf{s_0}$ (as processed from the raw $\mathbf{G}$ matrix from Section \ref{sec:system}\ref{ssec:setup}) is thus

\begin{figure}[t]
\centering
\includegraphics[width=3.4in]{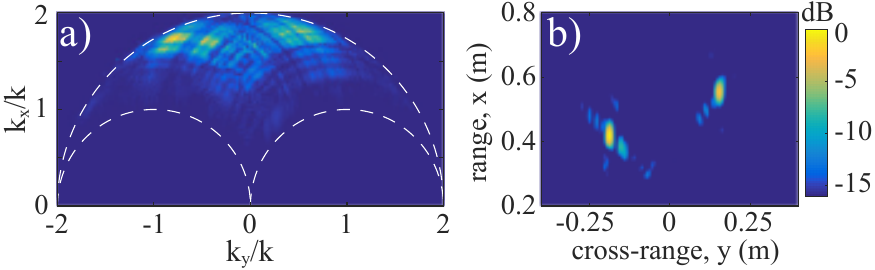}
\caption{Example of a) simulated data in $k$-space and b) its reconstruction which is found through the range migration algorithm. The white outline in (a) contains the portions of $k$-space that are probed by the system; the area outside represents non-propagating wave vectors.}
\label{fig:inv}
\end{figure}

\begin{equation}
\mathbf{s_0} = \bm{\Phi}_{Tx}^+ \mathbf{G} (\bm{\Phi}_{Rx}^+)^\text{T}
\label{eq:signaltrans}
\end{equation}

\noindent where

\begin{equation}
\bm{\Phi}_{Tx} = \mathbf{U}_{Tx} \bm{\Lambda}_{Tx} \mathbf{V}_{Tx}^*
\label{eq:sourcemat}
\end{equation}

\noindent is the SVD of the transmitter's source matrix and

\begin{equation}
\bm{\Phi}_{Tx}^+ = \mathbf{V}_{Tx} \bm{\Lambda}_{Tx}^+ \mathbf{U}_{Tx}^*
\label{eq:sourcematinv}
\end{equation}

\noindent is its pseudo-inverse. Within these equations $+$ stands for the pseudo-inverse operator, $*$ is the conjugate transpose operator, and $\text{T}$ stands for the transpose operator. The SVD entails splitting $\bm{\Phi}$ into two unitary matrices, $\mathbf{U}$ and $\mathbf{V}^*$, and a diagonal matrix, $\bm{\Lambda}$, which includes the singular values. All singular values falling below $1\%$ of the largest singular value are truncated. The same is done for the receiver.

In essence, the process in (\ref{eq:signaltrans}) mathematically reverts the SISO problem back to the simplest physical situation (the MIMO case, where each pair of positions on the Tx and Rx apertures are sampled separately \cite{Fromenteze2017}), but allows for a simpler hardware to carry out the process in a faster, more practical implementation \cite{Pulido-Mancera2016c}.

It should be noted that the SVD has long been used for image inversion and information theoretic techniques \cite{Piestun2000, Solimene2004, Marks2016, Gollub2017, Sleasman2016c}. Recent applications have studied system resolution based on the SVD \cite{Solimene2004} and have also studied how many independent measurements can be taken in an imaging system \cite{Piestun2000, Marks2016, Gollub2017, Sleasman2016c}. More commonly, the SVD has been used to invert matrix equation (\ref{eq:gHf}) and directly solve for $\hat{\bm{\sigma}}$ in the inverse scattering problem \cite{Piestun2000}. The $\mathbf{H}$ matrix, however, is significantly larger than $\bm{\Phi}$ so this approach becomes cumbersome when the problem is large. When the SVD is used for image inversion, the singular values are often truncated based on the signal-to-noise ratio (SNR) of the system. In our case, the SVD truncation is acting on $\bm{\Phi}$ which is not \emph{directly} related to the SNR of the system. Relating the truncation level to the system's noise characteristics will be a topic of future exploration in this algorithm.

We now briefly outline the reconstruction strategy employed in \cite{Fromenteze2017} which provides physical insight into the imaging process and highlights how the spatial diversity of the aperture enables monochromatic imaging in the range dimension. It should be kept in mind that the physical measurements will be obtained as in Fig. \ref{fig:concept}(c), as outlined in Section \ref{sec:system}, but that the mathematical framework detailed here is based on the depiction in Fig. \ref{fig:concept}(b).

From the SVD inversion we get the decoupled spatial contributions $s_0(y_t,y_r)$ along the aperture. These can be represented in the spatial frequency domain by taking the Fourier transform as 

\begin{equation}
S_0(k_{yt},k_{yr}) = \sum_{y_t} \sum_{y_r} s_0(y_t,y_r) e^{j(k_{yt} y_t + k_{yr} y_r)}.
\label{eq:ksignal}
\end{equation}

\noindent After completing phase compensation, invoking the stationary phase approximation, and accounting for the physically offset locations of the two apertures \cite{Pulido-Mancera2016c, Lopez-Sanchez2000, Zhuge2011} we can write this signal as

\begin{equation}
S_M(k_{yt},k_{yr}) = S_0(k_{yt},k_{yr}) e^{-j k_{yt} y_{t0}} e^{-j k_{yr} y_{r0}} e^{j k_{y} y_{C}} e^{j k_{x} x_{C}}
\label{eq:filtering}
\end{equation}

\noindent where $(x_C, y_C)$ is the center of the scene, $y_{t0}$ and $y_{r0}$ are the central locations of the transmitting and receiving apertures, and the $k$-components are defined in accordance with the free space dispersion relation. Specifically, the spatial frequency components combine as $k_x=k_{xt}+k_{xr}$ (and similarly for $k_y$) and $k_{xt}=\sqrt{k^2-k_{yt}^2}$ (and similarly for $k_{xr}$). Thus, the first two exponentials account for the locations of the apertures and the last two exponentials compensate to reference the phase with respect to the center of the scene.

Since $S_M(k_{yt},k_{yr})$ is defined on an irregular grid, it must be resampled onto a regular grid if we wish to use the traditional fast Fourier transform to reconstruct the spatial representation of the scene. Here we perform Stolt interpolation, a process that resamples our $k$-space data onto a uniform grid, and the regularly-sampled data is $S_I(k_x,k_y)$ \cite{Stolt1978, Zhuge2012, Fromenteze2017}. After we have completed our resampling, the image can be efficiently found as

\begin{equation}
\hat{\sigma}(x,y) = \mathcal{F}_{2D}^{-1} [S_I(k_x,k_y)].
\label{eq:inversion}
\end{equation}

The last two steps of this reconstruction process are shown in Fig. \ref{fig:inv}. The signal interpolated onto a regular grid, $S_I$, is shown in Fig. \ref{fig:inv}(a). Note that all signal should fall within the outlined region because the components outside are evanescent and will not contribute to imaging. For the signal in Fig. \ref{fig:inv}(a), generated from two scatterers in the scene, the inversion has been completed and the result is shown in Fig. \ref{fig:inv}(b). This method of reconstruction will be the primary method utilized in this work and it will be contrasted with the GMRES backprojection method in Section \ref{sec:exp}\ref{ssec:recon}.

The mapping from $S_M(k_{yt},k_{yr})$ to $S_I(k_x,k_y)$ is one of the more computationally expensive steps in the reconstruction process and can become cumbersome when the problem scales up or is extended to 3D. An alternative approach is to use the non-uniform fast Fourier transform (NUFFT)---to combine the resampling and FFT steps and complete the reconstruction directly \cite{dutt1993fast,LiuNUFFT}---or to implement lookup tables to streamline the interpolation process. These schemes are beyond the scope of the present paper and will be left for future work.

\section{Experimental Demonstration}
\label{sec:exp}

In this section we demonstrate the single-frequency imaging approach for a set of test targets, comparing experimentally-measured and analytically-predicted resolution in both range and cross-range. Then we investigate systematic misalignments and we contrast their impact on single-frequency imaging as compared to the finite bandwidth case. Finally, we compare the modified RMA with a least squares solver, GMRES, which solves (\ref{eq:gHf}). Through each of these studies, we comment on the impact of using only a single frequency versus a finite bandwidth.

\begin{figure}[t]
\centering
\includegraphics[width=3.3in]{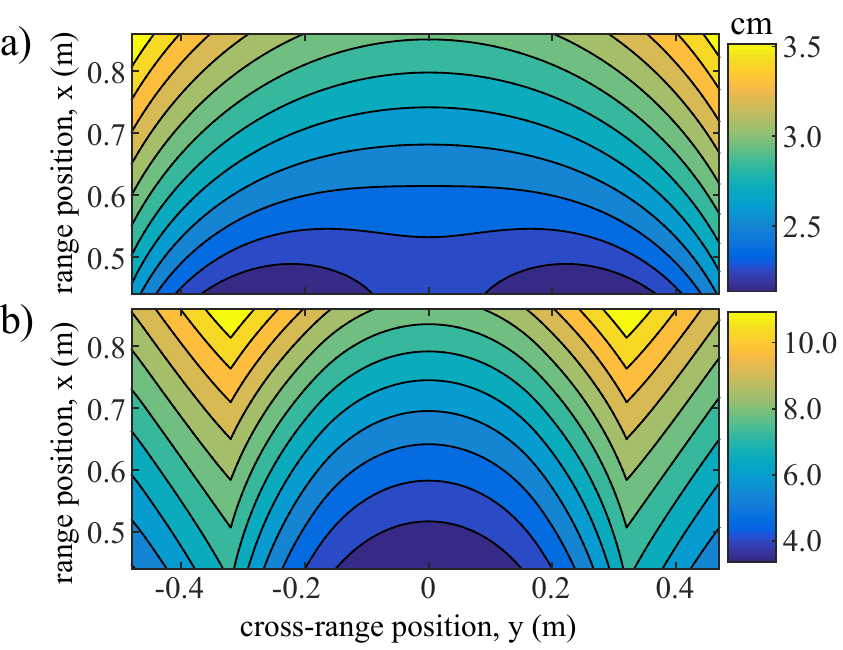}
\caption{Spatially varying a) cross-range resolution and b) range resolution, as calculated analytically from geometric factors including Eq. \ref{eq:kxt}.}
\label{fig:res}
\end{figure}

\subsection{Point Spread Function}
\label{ssec:PSF}

Based on the $k$-space analysis above, the illuminating fields should create sufficient support to resolve objects on the order of centimeters when located within \SI{1}{\meter} of the aperture. The $k$-space support can be determined analytically from the component-wise contributions of the transmitting and receiving apertures. Specifically, for the $x$ component and for the transmitter we can find

\begin{equation}
k_{xt} = \frac{2 \pi f}{c} \cos (\theta (\bar{r}_0-\bar{r}_t))
\label{eq:kxt}
\end{equation}

\noindent for $\bar{r}_t$ locations on the aperture and a location $\bar{r}_0$ in the scene (similarly for $k_{xr}$)---here $\theta$ is the angle of the vector between these two points, measured from the antenna's broadside. The combined spatial frequency in the $x$ direction from the transmitter and receiver is the same as the sum in (\ref{eq:kxSum}). A similar expression can be written for $k_{yt}$ and $k_{yr}$, with the cosine replaced by sine \cite{Ahmed2011}.

With all $k$-components calculated across the aperture for a specific location in the scene, the maximum/minimum $k_x$ and $k_y$ can be found. At the specific location $\bar{r}_0=(x_0, y_0)$ the best case range/cross-range ($x$/$y$) resolution is then determined according to (\ref{eq:delta_xy}). Calculating explicit equations for resolution in terms of scene location does not return a clean formula and is heavily dependent on the system geometry. For our setup (described in Section \ref{sec:system}\ref{ssec:recon}) the resulting resolutions have been plotted in Fig. \ref{fig:res}. The plotted resolution is determined from substituting $\Delta k_x$ and $\Delta k_y$, which are specifically based on the geometry of our system, into (\ref{eq:delta_xy}). It should be noted that this resolution is an optimistic upper bound since the $k$-space sampling is non-uniform and only has certain areas with sufficiently dense sampling.

\begin{figure}[t]
\centering
\includegraphics[width=3.3in]{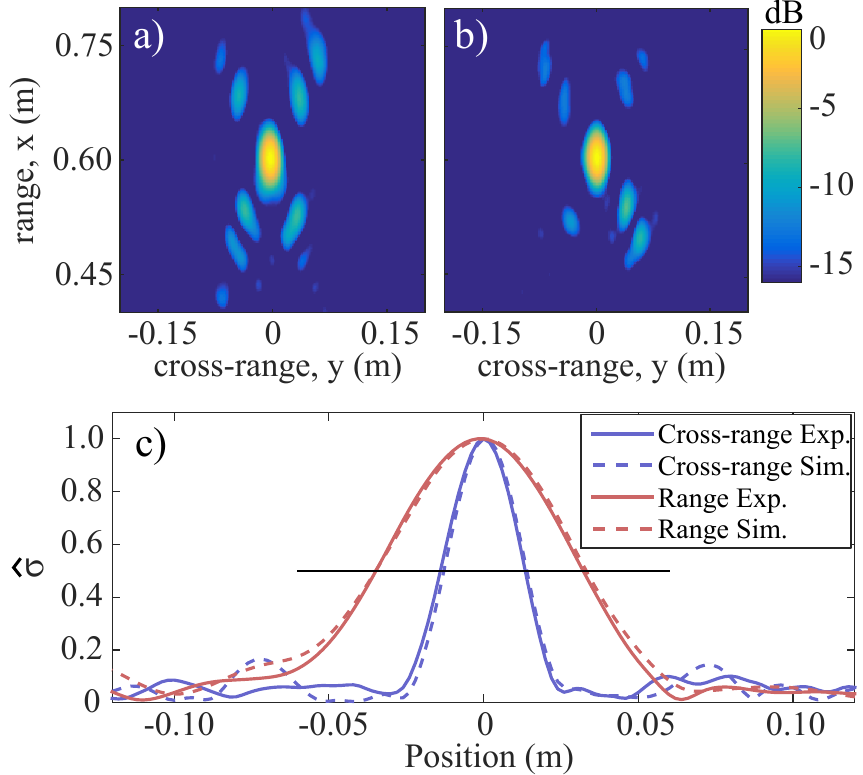}
\caption{a) Experimental PSF analysis, b) simulated PSF, and c) the cross sections along the main axes. All reconstructions are obtained with the RMA technique.}
\label{fig:psf}
\end{figure}

When a sub-diffraction-limited object is placed in the scene, we can experimentally estimate the resolution by taking the full width at half maximum (FWHM) of the resulting reconstructed image. The overall shape of this reconstruction will also identify the magnitudes and locations of the sidelobes, which would reveal any potential aliasing issues. This point spread function (PSF) analysis is thus an important characterization of the system's overall response \cite{Yurduseven2015,Soumekh1999}.

\begin{figure}[t]
\centering
\includegraphics[width=3.4in]{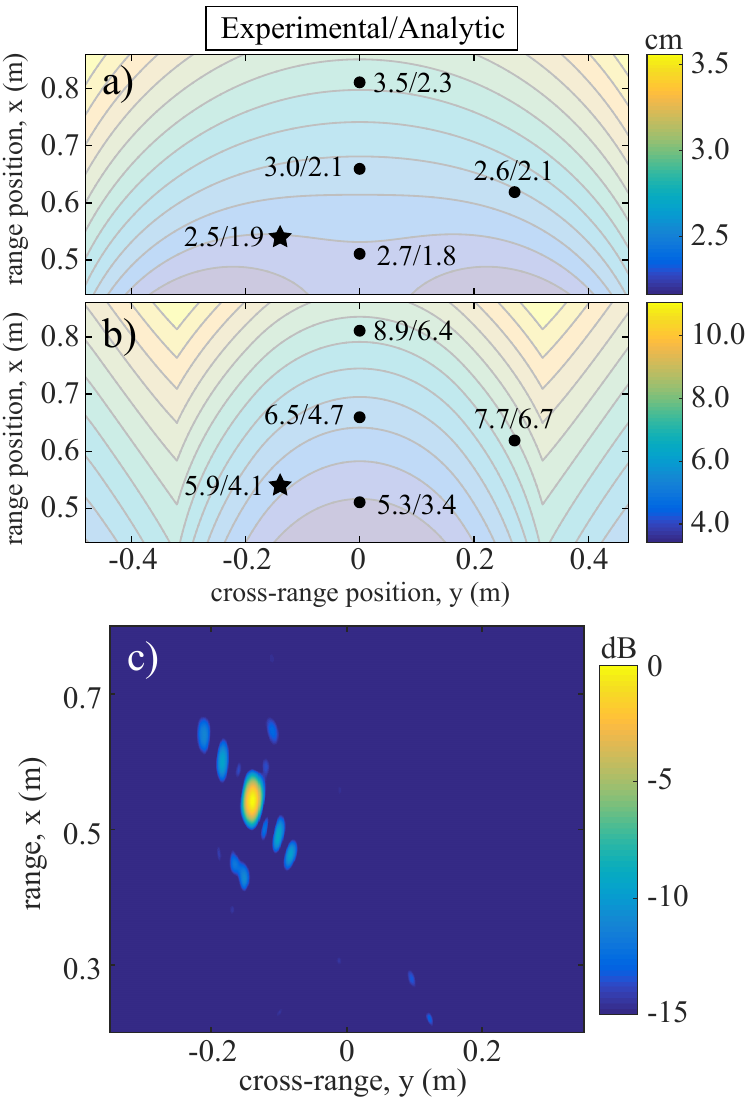}
\caption{Comparison of experimental data with analytic expectation for a) cross-range resolution and b) range resolution. c) An example of an off-centered reconstruction is also shown which corresponds to the case demarcated by a star.}
\label{fig:psf_varied}
\end{figure}

Figure \ref{fig:psf}(a) shows the experimental results of a PSF analysis with a metallic rod of diameter \SI{3}{\milli\meter} (a 2D point scatterer) placed at $x= \SI{0.5}{\meter}$, $y=\SI{0}{\meter}$. For this result we reconstruct with the RMA method detailed in Section \ref{sec:system}\ref{ssec:recon} and use $P=100$ tuning states to obtain a high-quality image. Slices of this image taken along the major axes---plotted in Fig. \ref{fig:psf}(c)---are used to determine the cross-range and range resolution. The same process is carried out for a simulation of the same configuration and these simulated results are also plotted. The simulation is conducted with the forward model in \cite{Lipworth2015}, which uses the near fields scans and propagates the sources to the objects with the free space Green's function---this is essentially the formulation in Eq. (\ref{eq:gHf}).

By comparing the simulated result to the experiment, it is seen that qualitatively similar behavior appears in both reconstructions. The PSF resembles the shape of an "X", due to the geometric placement of the antennas, which is seen in both reconstructions. Additionally, the 1D cross sections show that the expected resolution is met. For the cross-range resolution, the experiment results in \SI{2.7}{\centi\meter}, the numeric simulation returns \SI{2.7}{\centi\meter}, and the analytic expectation gives \SI{1.8}{\centi\meter}. Overall, the agreement between the three cases is reasonable. The experimental PSF shows excellent agreement with the simulated PSF and it also confirms the expectation that the analytic approach is overconfident. The range result reveals a similar trend. Experimental results show a resolution of \SI{5.3}{\centi\meter}, in contrast to an expectation of \SI{5.3}{\centi\meter} from simulation and \SI{3.4}{\centi\meter} from the analytic result. This good agreement confirms that the configuration matches the expected performance.

\begin{figure}[t]
	\centering
	\includegraphics[width=3.15in]{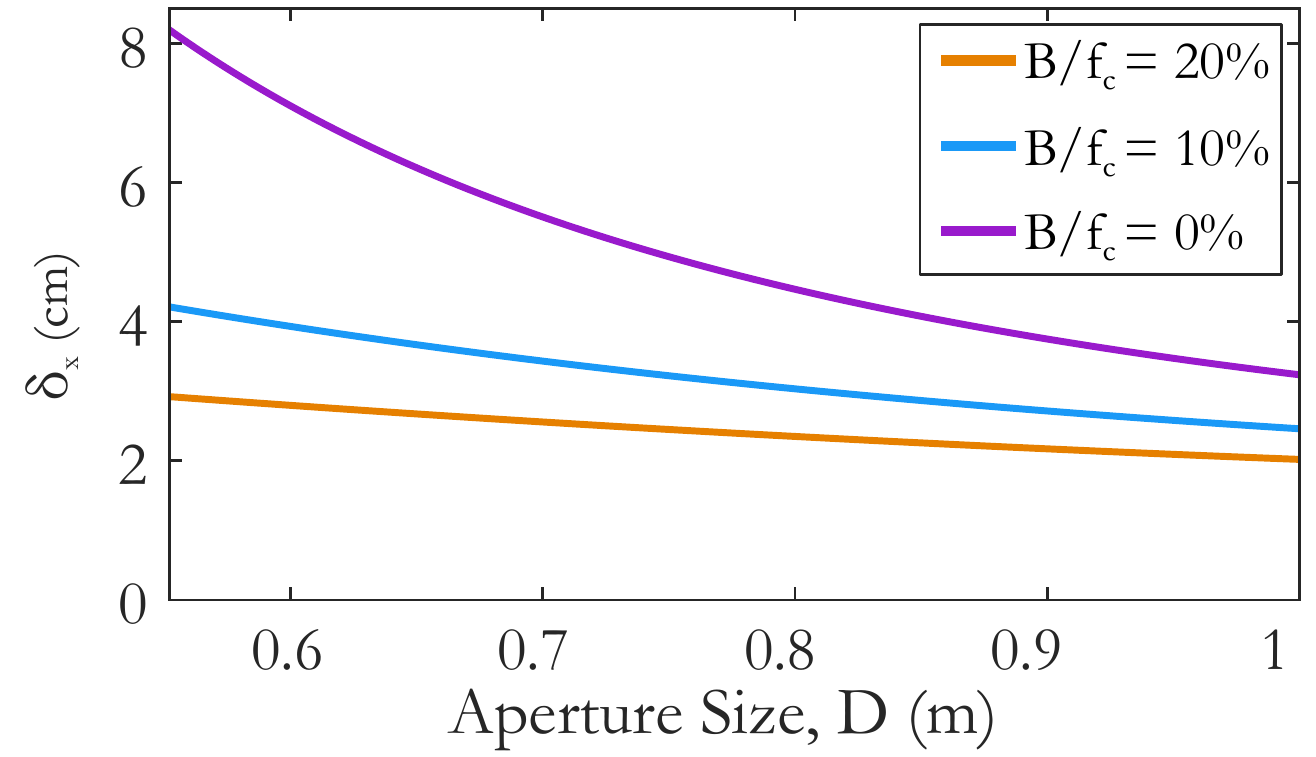}
	\caption{Resolution for a system with and without bandwidth. A hypothetical dynamic aperture is used as a transceiver with size $D$, bandwidth $B$, and central frequency $f_c$. The expected resolution is calculated at $(x=\SI{0.6}{\meter}, y=\SI{0}{\meter})$.}
	\label{fig:resolutionComp}
\end{figure}

The resolution of this imaging system varies across different locations in the region of interest. According to the analytic resolution plots in Fig. \ref{fig:res} we expect the cross-range and range resolutions to be spatially-variant. A PSF analysis is also conducted experimentally at several locations to demonstrate this effect. Figure \ref{fig:psf_varied} shows that the resolutions obtained from the reconstructed images are comparable to those from the analytic expressions. We also plot a reconstruction of an off-center object in Fig. \ref{fig:psf_varied}(c) as an example (corresponding to the star in Fig. \ref{fig:psf_varied}(a),(b)).

While the results presented in Figs. \ref{fig:res}-\ref{fig:psf_varied} establish that the proposed configuration is capable of resolving objects on the order of centimeters in both directions, it also raises an important question: how does the single-frequency imaging system compare to a traditional system which utilizes bandwidth? To answer this question, we examine a hypothetical situation where the same dynamic aperture, with length $D$, is used for both transmission and reception. In the same manner that Fig. \ref{fig:res} was generated (by calculating the $k$-space from the geometry) we can make this comparison when bandwidth is included. The range resolution along the $x$-axis can then be calculated as \cite{Ahmed2014}

\begin{equation}
\delta_x = \frac{c}{2} \Bigg( B+f_\text{min}\Bigg(1-\frac{1}{\sqrt{(1+(\frac{D}{2x})^2)}}\Bigg) \Bigg)^{-1}
\label{eq:123456}
\end{equation}

\noindent for the case of a transceiver (collocated Tx-Rx aperture) with size $D$ and bandwidth $B$. We have plotted curves for the case of $0\%$, $10\%$, and $20\%$ fractional bandwidth in Fig. \ref{fig:resolutionComp} for a point at $(x=\SI{0.6}{\meter}, y=\SI{0}{\meter})$. Although a larger bandwidth always enhances resolution, the resulting resolution with and without bandwidth are fairly comparable and converge as the aperture size grows. For a system defined as ultra wideband ($20\%$ fractional bandwidth) the resolution is better than the no bandwidth case ($0\%$) by a factor of $2{\times}$ when the aperture size is \SI{0.75}{\meter}, a feasible design for an implementation. Given the hardware benefits offered by operating at a single frequency, this factor of $2$ difference can be acceptable in many applications.

\subsection{Misalignment Sensitivity}

The aperture-to-aperture experiment here is an example of an electrically-large system with complex radiation patterns. In a recent work it was shown that such systems can suffer severely from misalignments between the antennas \cite{Odabasi2016}. Appreciable rotational or displacement errors can quickly degrade an image beyond recognition \cite{Steinberg1991}. This issue, for example, necessitated special efforts to ensure precise alignment, as outlined in \cite{Sleasman2017}, for the system examined in \cite{Gollub2017}. However, this sensitivity is exacerbated when a large bandwidth is used. Here we investigate the sensitivity of the single-frequency imaging system to misalignment and demonstrate that it is robust to practical misalignments.

\begin{figure}[t]
\centering

\begin{tikzpicture}

\node[anchor=south west,inner sep=0] (image) at (1.3,0) {\includegraphics[width=1.85in]{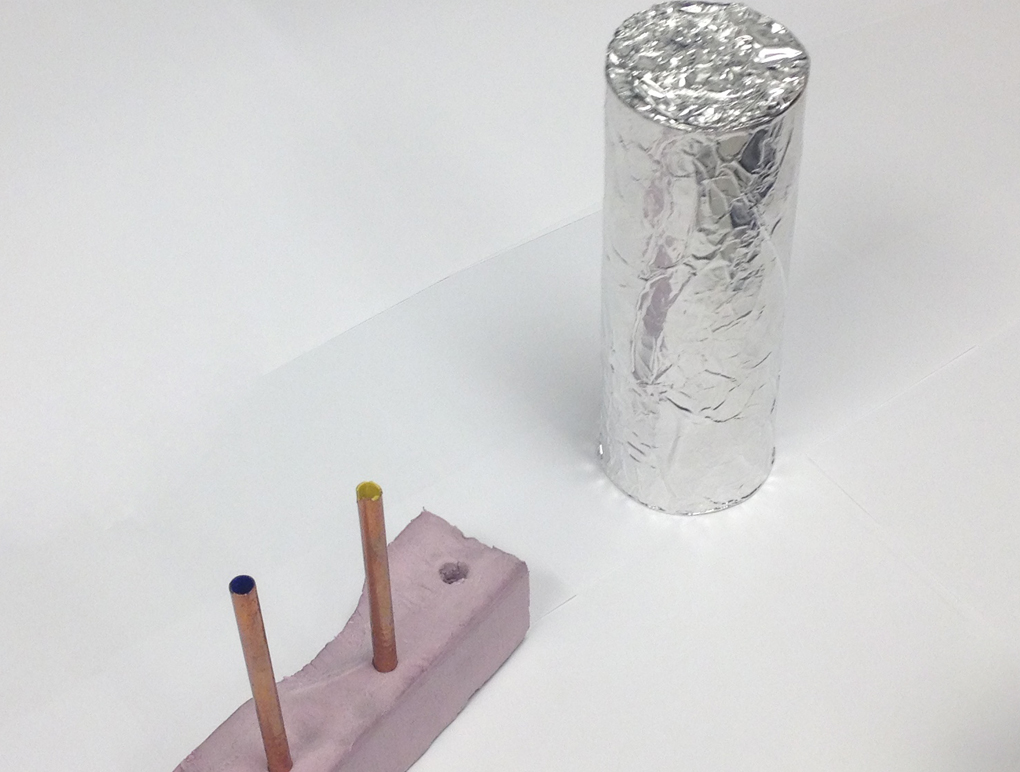}};

\node at (1.59,3.2) {\LARGE a)};

\node[anchor=south west,inner sep=0] (image) at (0,-8.78) {\includegraphics[width=3.0in]{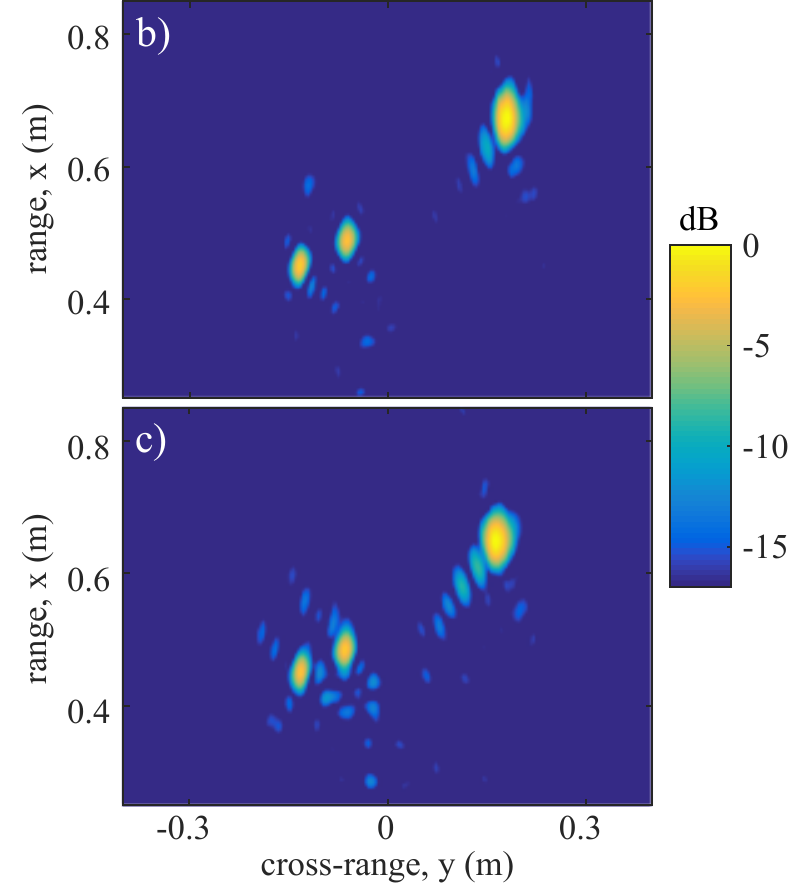}};

\end{tikzpicture}

\caption{Experimental alignment study with a) the scene composing three cylindrical objects. Reconstructions are completed with a) perfect alignment and b) intentional misalignment of \SI{-2.54}{\centi\meter} in the $x$ direction and \SI{-2.54}{\centi\meter} in the $y$ direction.}
\label{fig:misali_exp}
\end{figure}

We first obtain experimental images of objects when the antennas are ideally aligned, again with $P=100$ tuning states per antenna. After displacing the Tx antenna by a moderate amount ($\SI{-2.54}{\centi\meter}{=}1.6\lambda$ in $x$ and $\SI{-2.54}{\centi\meter}{=}1.6\lambda$ in $y$) we take new measurements and image assuming the location of the aperture has not been changed. The results for the aligned and misaligned cases are shown in Fig. \ref{fig:misali_exp}. As expected, misalignment has slightly degraded the quality of the image, but the image is still well-resolved and distortions are minor.

\begin{figure}[t]
\centering
\includegraphics[width=3.0in]{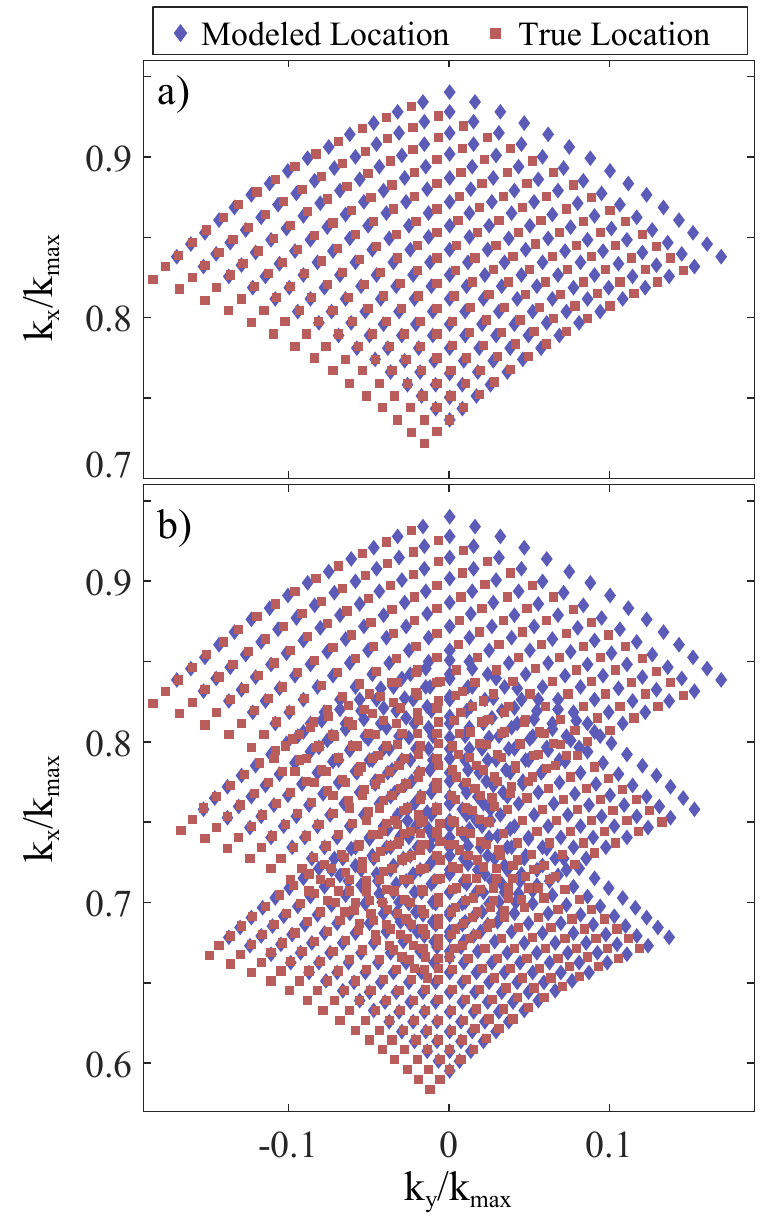}
\caption{$k$-space alignment effects for the cases of a) a single frequency and b) three frequencies across \SI{1.53}{\giga\hertz} of bandwidth.}
\label{fig:misali_k0+N}
\end{figure}

With a single frequency, the effects of misalignment in $k$-space are less pronounced than in wideband systems. Since the geometric factors alone determine the $k$-space sampling for the single frequency case, the result of a misalignment will be a smooth shifting with slight warping of the sampled $k$-components. Thus, except for extreme misalignments, we expect that the gradual shifting of the $k$-components will not significantly harm the imaging performance, but rather will lead to a minor displacement in the reconstructed objects' locations. Wideband systems, on the other hand, are more prone to degradation because of increased distortion in $k$-space. A physical interpretation of this effect would be that a given misalignment length has different electrical lengths at different portions of the frequency band of operation. Thus, the phase distortions are different for the different frequency points, leading to data that seems self-contradictory. We have plotted the sampled $k$-components for the single frequency case in Fig. \ref{fig:misali_k0+N}(a) and for a case with three frequencies in Fig. \ref{fig:misali_k0+N}(b)---the overlapping areas in the $B \neq \SI{0}{\giga\hertz}$ case lead to conflicting data which harms the reconstruction.

To show the effects of bandwidth and misalignment in imaging, we carry out simulations for the case of $B\,{=}\,0$ and $B\,{=}\,\SI{1.53}{\giga\hertz}$. When doing this we used GMRES and solve Eq. (\ref{eq:gHf}) so that the inversion in Eq. (\ref{eq:signaltrans}) will be fair across the simulations---this will become clearer in the next section. The same maximum frequency is used in both cases, with the $B\,{>}\,0$ simulation spanning \SIrange{17.5}{19.03}{\giga\hertz} with 18 points. To reduce the size of the problem (when $P\,{=}\,100$ with $18$ frequency points the $\mathbf{H}$ matrix becomes large) we use a smaller number of tuning states. It was found that $18$ frequencies combined with $8$ tuning patterns results in a reasonable image. To make the case with $B\,{=}\,0$ match, we reduce the number of tuning states to $P\,{=}\,34$. In this manner, both methods have approximately $1150$ measurements.

\begin{figure}[t]
\centering
\includegraphics[width=3.4in]{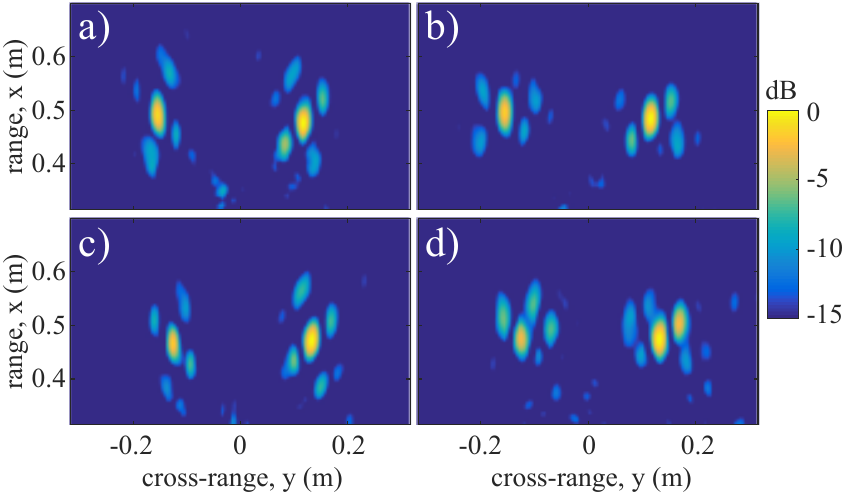}
\caption{Simulated results for a) $B=\SI{0}{\giga\hertz}$, perfect alignment b) $B=\SI{1.53}{\giga\hertz}$, perfect alignment, c) $B=\SI{0}{\giga\hertz}$, bad misalignment, and d) $B=\SI{1.53}{\giga\hertz}$, bad misalignment}
\label{fig:misali_sim}
\end{figure}

An image of two point scatterers is reconstructed with ideal alignment for both cases in Fig. \ref{fig:misali_sim}(a),(b). Misalignment of $x=\SI{2}{\centi\meter}$, $y=\SI{3}{\centi\meter}$ is then introduced (compared to a wavelength of $\lambda=\SI{1.6}{\centi\meter}$) and the simulation is repeated. The results for $B = \SI{0}{\giga\hertz}$ and $B=\SI{1.53}{\giga\hertz}$ are shown in Fig. \ref{fig:misali_sim}(c),(d) and it is clearly seen that the case with null bandwidth has greater fidelity. The case without bandwidth only exhibits a shift in the location and some minor distortion of the sidelobes. On the other hand, for the $B=\SI{1.53}{\giga\hertz}$ result the sidelobe of the object at $y=\SI{+0.2}{\centi\meter}$ is larger than the target itself at $y=\SI{-0.2}{\centi\meter}$. Furthermore, the object at $y=\SI{-0.2}{\centi\meter}$ has sidelobes that start to merge with the actual object.

The alignment sensitivity exemplifies yet another advantage of the single-frequency imaging paradigm. In addition, we note that the calibration procedure for the two cases also presents a contrasting feature. Calibration of a coherent system typically involves defining a reference plane (after the RF source) which the measurements can be taken with respect to. Specifically, it is essential that the phases of the different measurements can be considered together and compensated for. In the case of a single frequency, the reference plane can be chosen arbitrarily and all of the measurements will effectively have the same phase offset. For the case with multiple frequencies, each measurement will have a different phase offset at the reference plane and therefore it is necessary to correct for the differences. We do not experimentally demonstrate this notion here, but simply comment that this consideration becomes more important when the system scales up. For example, the imager in \cite{Gollub2017} has $24$ Rx ports and $72$ Tx ports, necessitating the development of special procedures to perform this phase calibration \cite{Yurduseven2016d}. A single-frequency system can operate without compensation so long as the radio has internal stability.

\subsection{GMRES vs. RMA Reconstructions}

Lastly, we demonstrate the fact that the RMA method has faster processing times and higher fidelity. To this end, we compare the performance of the RMA with GMRES (as implemented by the built-in MATLAB function). GMRES was completed with $1,643$ locations in the scene, with a ROI depth range of \SI{40}{\centi\meter} and a ROI width of \SI{68}{\centi\meter}, sampled on a $\SI{1.3}{\centi\meter}\times\SI{1.3}{\centi\meter}$ grid. The solver was run for $20$ iterations, taking a total time of \SI{1.46}{\second}. Pre-computation of the $\mathbf{H}$ matrix took a total of \SI{13.2}{\second}. On the other hand, the RMA reconstructs over a significantly larger area (defined by the sampling in $k$-space) in a shorter amount of time. The pre-computation (including SVD truncation of the source matrices) took \SI{0.22}{\second} and the direct inversion took \SI{0.16}{\second}. The direct computation time shows an order of magnitude speed enhancement. With the exception of the backprojection $\mathbf{H}$ matrix build, all of these computations were run on a CPU with a 3.50 GHz Intel Xeon processor and 64 GB of RAM. The computation of the $\mathbf{H}$ matrix was completed on a NVIDIA Quadro K5000 GPU. All of these computations can generally be implemented on a GPU which will provide further time advantages, but the GMRES is inherently iterative and therefore does not receive as large a benefit from this transition. Further research in this area is a topic of much interest and optimized reconstruction methods geared around single-frequency imaging will be investigated in future works.

\begin{figure}[t]
\centering
\begin{tikzpicture}

\node[anchor=south west,inner sep=0] (image) at (1.05,0) {\includegraphics[width=2.10in]{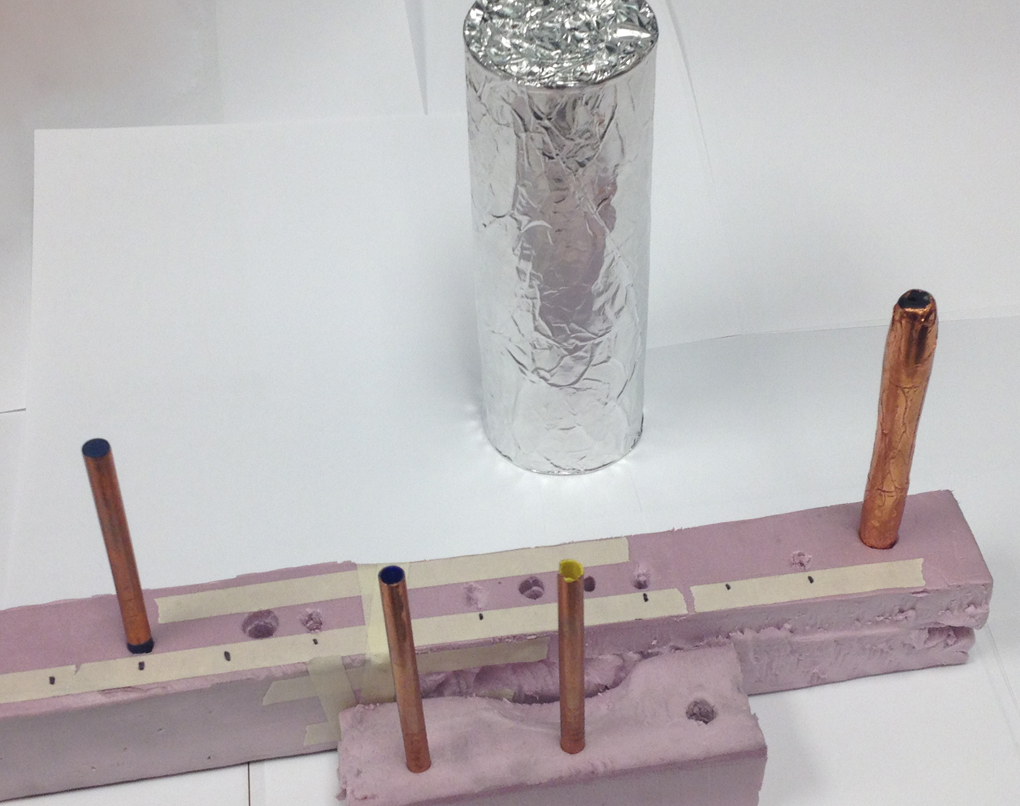}};

\node at (1.3,3.88) {\LARGE a)};

\node[anchor=south west,inner sep=0] (image) at (0,-7.31) {\includegraphics[width=3.0in]{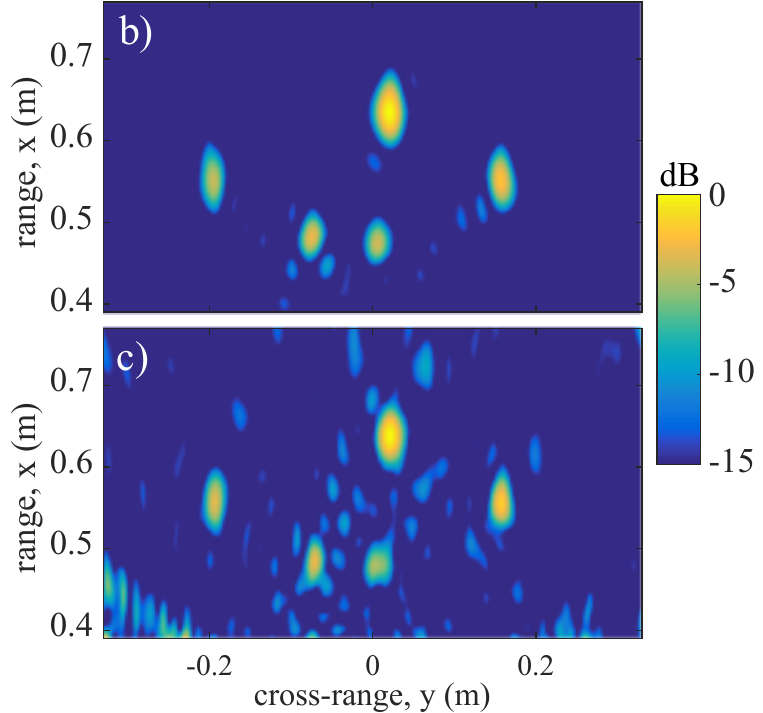}};

\end{tikzpicture}
\caption{Experimental comparison of a) the scene objects for the cases of b) RMA and c) GMRES.}
\label{fig:recon}
\end{figure}

A contrast between the two methods, GMRES and RMA, lies in how the source distribution across the aperture $\bm{\Phi}$ is taken into consideration. In the implementation based on Eq. (\ref{eq:gHf}), the sources representing the dynamic metasurface aperture are considered directly in the forward propagation, whereas in the RMA method a pre-processing inversion step is required (completed for each frequency if bandwidth is used). When the number of tuning states is low ($P\,{<}\,20$) the inversion in (\ref{eq:signaltrans}) becomes less accurate and the subsequent reconstruction quality degrades. Since GMRES acts on the total $\mathbf{H}$ matrix, the number of frequencies versus number of tuning states is irrelevant to its operation; thus it was the method of choice for comparing the misalignment simulations which had a variable number of frequencies (and thus a variable $P$). A more detailed study on the effects of SVD truncation and inversion can be found in \cite{Pulido-Mancera2016c}.

The result that we emphasize here is the superior reconstruction quality of the RMA in Fig. \ref{fig:recon} and the shorter processing time. For this reconstruction, $P\,{=}\,100$ tuning states were used and a complex scene, shown in Fig. \ref{fig:recon}, was imaged. It is worth noting that the RMA reconstructs a significantly larger ROI (over \SI{1}{\meter} in both range and cross-range) and that we have zoomed-in on the targets after reconstructing.

\section{Conclusion and Discussion}
\label{sec:conclude}

We have shown that the fundamental idea of range-imaging with a single frequency is tenable with the appropriate hardware. Additionally, a dynamic metasurface aperture implementation was demonstrated that can conduct single-frequency computational imaging from a low-profile and cost-effective platform. The dynamic aperture can be rapidly reconfigured to multiplex the information in a scene with no bandwidth.

The numerous advantages of single-frequency imaging have been highlighted and demonstrated. Specifically, this architecture is attractive as a favorable hardware layer capable of obtaining high-fidelity images in the Fresnel zone. For applications involving dispersive objects, this platform may find particular advantage because the single-frequency measurements are not be sensitive to dispersion. Additionally, calibration of the feed network is not required and the system is highly robust to misalignment errors. Lastly, removing the requirement for a large bandwidth means that many portions of the electromagnetic spectrum will be available and that interference from other devices may be less pronounced. The high-quality performance and unique advantages of this system pose it as a powerful architecture for next generation microwave computational imaging applications.

\section{Funding Information}
This work was supported by the Air Force Office of Scientific Research (AFOSR, Grant No. FA9550-12-1-0491).

\bibliography{main}

\end{document}